\documentclass[letterpaper,aps,prd,reprint,floatfix,amsmath,amssymb]{revtex4-1}
\usepackage{graphicx}
\usepackage{color}
\usepackage{dcolumn}\usepackage{bm}
\newcommand{\ud}{\,\mathrm{d}}
\newcommand{\ue}{\,\mathrm{e}}

\begin{document}

\title{Quantizing the Homogeneous Linear Perturbations about Taub using the Jacobi Method of Second Variation}
\author{Joseph H. Bae}
\affiliation{Department of Physics, Yale University, New Haven, CT 06520, USA.}
\date{\today}
\begin{abstract}
Applying the Jacobi method of second variation to the Bianchi IX system in Misner variables ($\alpha, \beta_+, \beta_-$), we specialize to the Taub space background ($\beta_-=0$) and obtain the governing equations for linearized homogeneous perturbations ($\alpha', \beta_+', \beta_-'$) thereabout. Employing a canonical transformation, we isolate two decoupled gauge-invariant linearized variables ($\beta_-'$ and $Q_+' = p_+\alpha'+p_\alpha \beta_+'$), together with their conjugate momenta and linearized Hamiltonians. These two linearized Hamiltonians are of time-dependent harmonic oscillator form, and we quantize them to get time-dependent Schr\"{o}dinger equations. For the case of $Q_+'$, we are able to solve for the discrete solutions and the exact quantum squeezed states. 
\end{abstract} 
\maketitle

\section*{Introduction}
The Taub space is a special case of Bianchi IX; unlike Bianchi IX, whose exact solutions are known only asymptotically, the Taub metric is an exact solution to Einstein's equations \cite{Taub51, Latifi}. In this paper, we use the Jacobi method of second variation to study the homogeneous linearized perturbations about the Taub background solutions (For the method applied to the Schwarzschild metric, see \cite{Moncrief73}). Because the Taub background solutions are known exactly, we can write down the explicit solutions of the linearized Hamilton equations for the linear perturbations that stay within the Taub model. These explicit solutions, together with the method of invariants \cite{LewisR}, allow us to solve for the quantum squeezed states  for the quantized linearized Hamiltonians.

The paper is organized as follows: we start with ADM action of the background Bianchi IX expressed in Misner variables, paying special attention to the background Hamiltonian constraint. Next, we apply the Jacobi method of second variation, obtaining the linearized constraint and the linearized Hamiltonian for the perturbation variables. A canonical transformation decouples the linearized Hamiltonian into gauge-invariant and gauge-dependent parts. Of the gauge-invariant part of the linearized Hamiltonian, we identify the part that stays within the Taub family of solutions, writing down the explicit time-dependences of the new gauge-invariant perturbation variable from the exact solutions for the background Taub space. Finally, we use the method of invariants to calculate the quantum squeezed states from the exact classical solutions.

\section{Bianchi IX using Miser Variables}
We can express the dimensionless line element of Bianchi IX model using Misner variables $\{\alpha,\beta_+, \beta_+\}$. Setting $\beta_- = 0$ would make this into a Taub model. We have divided the line element by $r^2 = \frac{l_0{}^2}{6\pi}$, and defined a dimensionless time parameter $\tau = \frac{t}{r}$:
\begin{equation}
\begin{split}
\ud& s^2 = - N^2 \ud \tau^2\\&  + \begin{pmatrix} \ue^{2\alpha + 2\beta_++2\sqrt{3}\beta_-}&0&0\\0&\ue^{2\alpha + 2\beta_+-2\sqrt{3}\beta_-}&0\\0&0&\ue^{2\alpha -4\beta_+} \end{pmatrix}_{\!\!ij} \sigma^i \sigma^j.\label{MisnerMetric}
\end{split}
\end{equation}
where $\sigma^i$'s are the invariant differential one-forms on the $\mathbb{S}^3$ manifold satisfying $\ud \sigma^i = \frac{1}{2}\epsilon_{ijk} \, \sigma^j \wedge \sigma^k$. Following the method outlined in \cite{Bae2014}, we write the ADM action of Bianchi IX:
\begin{equation}
S_{ADM} = \int \Big( p_\alpha \dot{\alpha} + p_+ \dot{\beta_+}  + p_- \dot{\beta_-}  - N H_\perp  \Big) \ud \tau.
\end{equation}
Variation of the lapse function $N$ leads to the Hamiltonian constraint,
\begin{widetext}
\begin{equation}
H_\perp  = - \rho \ue^{-3\alpha} \bigg[ \gamma_- - \tfrac{\lambda}{\rho} \ue^{4\alpha} \Big(\ue^{-8\beta_+}  - 4 \ue^{-2\beta_+}\cosh{2\sqrt{3}\beta_-}  + 2\ue^{4\beta_+}\big( \cosh{4\sqrt{3}\beta_-}-1\big) \Big) \bigg] = 0.\label{Hconstraint}
\end{equation}
\end{widetext}
We introduce the following abbreviations used throughout this paper:
\begin{eqnarray}
\rho  =& \frac{G}{24\pi r^3}, &\\ \lambda  =& \frac{\pi r}{2 G}, &\\
 \gamma_\pm  = &\pm p_\alpha{}^2 \mp p_+{}^2 \mp p_-{}^2.&
\end{eqnarray}

\section{Jacobi Method of Second Variation}

We apply the Jacobi method of second variation to write the linearized Hamiltonian for homogeneous perturbations about the background Taub solution (for a similar approach applied to the Schwarzschild metric, see \cite{Moncrief73}). This is equivalent to a linear approximation about the Taub solution (denoted by `${}_0$' suffix):
\begin{eqnarray}
\alpha &= \alpha_0 + \alpha' \\
\beta_\pm & = \beta_\pm{}_0 + \beta_\pm' \\
p_\alpha & = p_\alpha{}_0 + p_\alpha' \\
p_\pm & = p_\pm{}_0 + p_\pm' \\
N & = N_0 + N'
\end{eqnarray}
When applied to our ADM Hamiltonian formulation, the approximation yields a linearized constraint (we drop the `${}_0$' suffix for clarity) and a linearized Hamiltonian for the variations. In effect, we obtain another variational principle whose Hamilton equations govern the homogeneous perturbations about the exact Taub solution.

Another way to think about the linear approximation is to consider each `background' variable as a function of an extra parameter (say $e$), and define the `primed' notation as:
\begin{equation}
q'= \frac{\partial q(\tau,e)}{\partial e}\bigg|_{e=0}. \label{primedpara}
\end{equation}
This is exactly what is done later in Eq.~(\ref{q1para}) and (\ref{q2para}).
Write the resulting linearized action:
\begin{equation}
S_{ADM}'' = 2 \int \Big( p_\alpha' \dot{\alpha'} + p_+' \dot{\beta_+'}  + p_-' \dot{\beta_-'}  - H_{}''  \Big) \ud \tau,
\end{equation}
where the linearized constraint and the linearized Hamiltonian are: 
\begin{eqnarray}
H_{}'' = N' H_\perp' + \tfrac{1}{2}N H_\perp '', \label{Hbreakdown}\\
H_\perp' = - \rho \ue^{-3\alpha} \Big[ \gamma_-' - 4 \tfrac{\lambda}{\rho} \ue^{4\alpha} f_A \Big( \alpha'-\tfrac{2f_B}{f_A}\beta_+' \Big) \Big],\label{LConstraint}
\end{eqnarray}
\begin{widetext}
\begin{equation}
H_\perp'' = 2 \rho \ue^{-3\alpha} \bigg[ -p_\alpha'{}^2 +p_+'{}^2 + p_-'{}^2 + 3 \gamma_-'  \alpha'  
+ 4\gamma_- \Big( -(\alpha')^2-\tfrac{2f_B}{f_A}(\alpha')(\beta_+') 
+ \tfrac{2f_C}{f_A}(\beta_+')^2 - 6\tfrac{f_D}{f_A}(\beta_-')^2 \Big) \bigg] .
\end{equation}
\end{widetext}
Variation of the linearized lapse function $N'$ leads to the linearized constraint $H_\perp ' = 0$. In the above equations and throughout this paper, the following abbreviations are used:
\begin{eqnarray}
f_A = \ue^{-8\beta_+} - 4\ue^{-2\beta_+}\\
f_B = \ue^{-8\beta_+} - \ue^{-2\beta_+}\\
f_C = 4\ue^{-8\beta_+} - \ue^{-2\beta_+}\\
f_D = \ue^{-2\beta_+} - 2\ue^{4\beta_+}.
\end{eqnarray}
Note that the linearized action does not involve the terms $(N'',\alpha'', \beta_+'', \beta_-'')$ or their conjugate momenta, as these are multiplied by terms that vanish identically when the background constraint and equations of motions are taken into account. For example, the term $N''$ would be multiplied by the background Hamiltonian $H_\perp$, which is zero by Eq.~(\ref{Hconstraint}). Note also that the linearized constraint Eq.~(\ref{LConstraint}) involves $(\alpha', \beta_+')$ and their conjugate momenta only. 

\subsection{Decoupling the linearized Hamiltonian}
The linearized Hamiltonian $H_{\perp}''$, together with the linearized constraint $H_\perp'=0$, govern the equations of motion for the perturbations. $H_{\perp}''$ naturally decouples into two Hamiltonians:
\begin{eqnarray}
H'' |_{H_\perp' = 0} = \tfrac{1}{2}N H_{\perp}'' = H_A'' + H_B'',\\
\begin{split}
H_A'' = \rho N \ue^{-3\alpha} \bigg[ -(p_\alpha')^2 +(p_+')^2+ 3 \gamma_-'  \alpha' \quad\\+ 4\gamma_- \Big( -(\alpha')^2-\tfrac{2f_B}{f_A}(\alpha')(\beta_+') + \tfrac{2f_C}{f_A}(\beta_+')^2 \Big) \bigg], \end{split}\\
H_B'' = \rho N \ue^{-3\alpha} \bigg[ (p_-')^2 -24 \gamma_- \tfrac{f_D}{f_A}(\beta_-')^2      \bigg]. \label{HB}
\end{eqnarray}
Because $H_B''$ involves $\beta_-'$ and $p_-'$ only, it is gauge-invariant: the equations of motion for these quantities do not depend on our choice of the linearized lapse function $N'$. Furthermore, $H_A''$ itself can be decoupled into a gauge-invariant part and a gauge-dependent part through a canonical transformation from $q_i' = \{\alpha', \beta_+'\}$ to $Q_i' = \{Q_+', Q_-'\}$ such that one of the two new conjugate momenta ($P_-'$) is proportional to $H_\perp'$: 
\begin{eqnarray}
Q'_i = \Lambda^j_i q'_j\\
P'^i = p'^j \Lambda^{-1} {}^i _j - \tfrac{\partial F}{\partial Q'_i}\\
\Lambda  = \begin{pmatrix} p_+ & p_\alpha \\-p_\alpha  & p_+\end{pmatrix}
\end{eqnarray}
\begin{widetext}
\begin{equation}
K_{A}'' = H_{A}''(P', Q') + \tfrac{\partial F}{\partial t} 
- (P'^j + \tfrac{\partial F}{\partial Q'_j})\Lambda^i_j (\dot{\Lambda}^{-1})^k_i Q'_k
\end{equation}
\begin{equation}
F  = \tfrac{1}{\gamma_+}\Big(  \tfrac{2 p_+ p_\alpha}{\gamma_+} (p_+ - \tfrac{2f_B}{f_A}p_\alpha) + (p_\alpha - \tfrac{2f_B}{f_A}p_+ ) \Big) Q_+'{}^2
 - \tfrac{2\gamma_-}{\gamma_+{}^2}(p_+ -\tfrac{2f_B}{f_A}p_\alpha)Q_+'Q_-'
   + \tfrac{\gamma_-}{\gamma_+{}^2}(p_\alpha + \tfrac{2f_B}{f_A}p_+)Q'_-{}^2
\end{equation}
Let us write out $K_{A}'' $ in full. Note that all the parts multiplied by $P'_-\propto H_\perp'$ go to zero when we impose the linearized constraint $H_\perp' = 0$.
\begin{equation}
\begin{split}
K_{A}''  = \rho N \ue^{-3\alpha} \bigg[ \gamma_- P_+'{}^2 - \tfrac{72}{f_A^2}\ue^{-10\beta_+}Q_+'{}^2  \bigg] \\
+ P_-' \rho N \ue^{-3\alpha}  \bigg[ \tfrac{2}{\gamma_+}(5p_\alpha{}^2-3p_+{}^2-8\tfrac{2f_B}{f_A}p_+p_\alpha)p_+Q_+'  
+\tfrac{2}{\gamma_+}(7p_\alpha p_+{}^2 - p_\alpha{}^3 + 4\tfrac{2f_B}{f_A}\gamma_-p_+)Q_-'\\
+\gamma_+ (\tfrac{2N'}{N} + 2\tfrac{2p_+p_\alpha}{\gamma_+}P_+' - \tfrac{\gamma_-}{\gamma_+} P_-')       \bigg]
\label{Ksuper}
\end{split}
\end{equation}
Thus, we have decoupled the two gauge-invariant linearized Hamiltonians for $\beta_-'$ and $Q_+'=p_+\alpha' + p_\alpha \beta_+'$ (Eq.~(\ref{HB}) and Eq.~(\ref{Ksuper}) respectively). We now proceed to solve the linearized Hamiltonian for $Q_+'$ (Eq.~(\ref{Ksuper})) explicitly. To do so, we will first need to use an explicit solution to the background Taub family.

\subsection{Exact Background Taub}
The solution for the Taub metric as given in A. Taub's 1951 paper \cite{Taub51} uses a notation and gauge that are similar to those used by Landau \& Lifshitz \cite{LandauL} as well as others \cite{GibbonsPage, Latifi} in describing the diagonalized Bianchi IX metric. Write the dimensionless line element (after dividing by $r^2$):
\begin{equation}
\ud s^2 = - (abc)^2 \ud \tau^2 + \begin{pmatrix} a^2&0&0\\0&b^2&0\\0&0&c^2 \end{pmatrix}_{\!\!ij} \sigma^i \sigma^j.
\end{equation}
The solutions are as follows:
\begin{eqnarray}
a^2 =  &\dfrac{2k}{\cosh{[2k(\tau-\tau_1)]}}\\
b^2 = c^2  = &\dfrac{k\cosh{[2k(\tau-\tau_1)]}}{2(\cosh{[k(\tau-\tau_2)]})^2}\\
N_I = &\sqrt{\dfrac{k^3 \cosh{[2k(\tau-\tau_1)]}}{2(\cosh{[k(\tau-\tau_2)]})^4}}.
\end{eqnarray}
Here $k$, $\tau_1$, and $\tau_2$ are arbitrary, real parameters; $\tau$ could be shifted to absorb $\tau_2$, leaving two independent dimensionless parameters $k$ and $\tau_1$ (We set $\tau_2 =0$ from here on.). In this gauge ($N_I=\ue^{3\alpha}$), the Misner variables $\alpha$ and $\beta_+$ have the following $\tau$-dependence:
\begin{eqnarray}
\ue^{6\alpha} =& \dfrac{k^3\cosh{[2k(\tau-\tau_1)]}}{2(\cosh{k\tau})^4}\label{6alpha} \\
\ue^{3\beta_+} =&\dfrac{\cosh{[2k(\tau-\tau_1)]}}{2\cosh{k\tau}} \label{6beta}\\
\ue^{2\alpha - \beta_+} = & \dfrac{k}{\cosh{k\tau }}.
\end{eqnarray}

\section{$Q_+'$ Hamiltonian}

Now let us write the linearized Hamiltonian with all the explicit $\tau$-dependences:
\begin{equation}
K_{A}'' = \rho \Big(p_\alpha{}^2 - p_+{}^2 \Big) P_+'{}^2 
-18 \rho \bigg( \frac{\cosh{k\tau} \cosh{[2k(\tau-\tau_1)]}}{\cosh{^2k\tau} - \cosh{^2[2k(\tau-\tau_1)]}} \bigg)^2 Q_+'{}2
\end{equation}
Comparing with the standard time-dependent harmonic oscillator equation, we arrive at
\begin{eqnarray}
K_{A}'' = \frac{P_+'{}^2}{2M(\tau)} + \frac{1}{2}M(\tau) \omega^2(\tau) Q_+'{}^2\\
M = \frac{-6 \rho}{k^2\Big[ \tanh{^2[2k(\tau-\tau_1)]} - \tanh{^2k\tau}\Big]}\\
\omega^2 = - 6 k^2 \Big[ \tanh{^2[2k(\tau-\tau_1)]} - \tanh{^2k\tau}\Big] 
\bigg(\frac{\cosh{k\tau} \cosh{[2k(\tau-\tau_1)]}}{\cosh{^2k\tau} - \cosh{^2[2k(\tau-\tau_1)]}} \bigg)^2
\end{eqnarray}
Following \cite{Dantas92}, we transform this to $H_1(\tau)$,
\begin{equation}
H_1 = \frac{p'{}^2}{2m} + \frac{1}{2}m \Omega^2(\tau) q'{}^2 \label{H1} \\
\end{equation}
where 
\begin{equation}
\begin{split}
\Omega^2 = \omega^2 + \frac{1}{4}\bigg( \frac{\dot{M}}{M}\bigg)^2 - \frac{1}{2}\bigg( \frac{\ddot{M}}{M}\bigg)\\
\Omega^2= -2 k^2\bigg[ \frac{4 \cosh{^4 k \tau}+ \cosh{^4 [2k(\tau-\tau_1)]}}{\cosh{^2[2k(\tau-\tau_1)]}- \cosh{^2 k\tau}}
+ \frac{-6\cosh{^2 k\tau}- 9\cosh{^2 [2k(\tau-\tau_1)]}}{\cosh{^2[2k(\tau-\tau_1)]}- \cosh{^2 k\tau}}\\
+ \frac{10 \cosh{^2 k\tau} \cosh{^2 [2k(\tau-\tau_1)]} }{\cosh{^2[2k(\tau-\tau_1)]}- \cosh{^2 k\tau}}
+ \frac{ - 3 \sinh{2k \tau }\sinh{[4k(\tau-\tau_1)]}}{\cosh{^2[2k(\tau-\tau_1)]}- \cosh{^2 k\tau}}\bigg],
\end{split}
\end{equation}
and the new canonical variables are related to the old by:
\begin{eqnarray}
q' = \bigg( \frac{M(\tau)}{m}\bigg)^{\frac{1}{2}} Q_+' = \bigg( \frac{M(\tau)}{m}\bigg)^{\frac{1}{2}} \frac{1}{2\rho} \bigg[ \frac{\partial \beta_+}{\partial \tau} \alpha' - \frac{\partial \alpha}{\partial \tau} \beta_+'  \bigg] \\
p' = \bigg( \frac{m}{M(\tau)}\bigg)^{\frac{1}{2}} P_+' + \frac{1}{2}\Big( m M(\tau) \Big)^{\frac{1}{2}}\frac{\dot{M}}{M} Q_+'
\end{eqnarray}
Now, from our explicit equations for $\alpha$ and $\beta_+$ as given in Eq.~(\ref{6alpha}-\ref{6beta}), we can write down the linear perturbation by simply differentiating $\alpha$ and $\beta_+$ with respect to the two parameters $\tau_1$ and $k$ (recall our definition earlier in Eq.~(\ref{primedpara})). We denote them by $q_1$ and $q_2$ respectively:
\begin{eqnarray}
q_1  = q' \bigg|_{q_i' = \frac{\partial q_i}{\partial \tau_1}}= \bigg( \frac{M}{m}\bigg)^{\frac{1}{2}}\frac{1}{2\rho} \bigg[\frac{\partial \beta_+ }{\partial \tau}\frac{\partial \alpha}{\partial \tau_1} - \frac{\partial \alpha}{\partial \tau} \frac{\partial \beta_+}{\partial \tau_1} \bigg] 
 = i \frac{k}{\sqrt{6\rho}} \frac{\tanh{k \tau} \tanh{[2k(\tau-\tau_1)]}}{\big(\tanh{^2[2k(\tau-\tau_1)]}-\tanh{^2k\tau}\big)^{\frac{1}{2}}} \label{q1para} \\
q_2 = q' \bigg|_{q_i' = \frac{\partial q_i}{\partial k}} = \bigg( \frac{M}{m}\bigg)^{\frac{1}{2}}\frac{1}{2\rho} \bigg[\frac{\partial \beta_+ }{\partial \tau}\frac{\partial \alpha}{\partial k} - \frac{\partial \alpha}{\partial \tau} \frac{\partial \beta_+}{\partial k} \bigg] 
 = i \frac{1}{2k \sqrt{6\rho}} \frac{\tanh{k \tau}- 2 \tanh{[2k(\tau-\tau_1)]}}{\big(\tanh{^2[2k(\tau-\tau_1)]}-\tanh{^2k\tau}\big)^{\frac{1}{2}}}
+ \frac{\tau_1}{k}q_1. \label{q2para}
\end{eqnarray}
\end{widetext}
From these two, we can construct two independent functions $\alpha_1$ and $\alpha_2$:
\begin{eqnarray}
\alpha_1 = \frac{\tanh{k \tau} \tanh{[2k(\tau-\tau_1)]}}{(\tanh{^2[2k(\tau-\tau_1)]}-\tanh{^2k\tau})^{\frac{1}{2}}}\\
\alpha_2 = \frac{1}{2}\frac{\tanh{k \tau}- 2 \tanh{[2k(\tau-\tau_1)]}}{(\tanh{^2[2k(\tau-\tau_1)]}-\tanh{^2k\tau})^{\frac{1}{2}}}.
\end{eqnarray}
By construction, then, $\alpha_i$ are solutions to the second order Euler-Lagrange equation that results from the time-dependent harmonic oscillator Hamiltonian Eq.~(\ref{H1}):
\begin{equation}
\ddot{\alpha}_i + \Omega^2 \alpha_i = 0.
\end{equation}
We also note that they satisfy the following initial conditions:
\begin{eqnarray}
\alpha_1(0) = 0\\
\dot{\alpha}_1(0)=- k \,\mathrm{sign}{(k\tau_1)} \\
\alpha_2(0) = \mathrm{sign}{(k\tau_1)}\\
\dot{\alpha}_2(0) = \frac{k}{ 2| \tanh{2k\tau_1}| }
\end{eqnarray}

\subsection{Discrete Solutions and Squeezed States}
Now, we are in a position to write down the discrete solutions and squeezed state solutions to the time-dependent Schr\"{o}dinger's equation:
\begin{equation}
H_1 \Psi(q',\tau) = - \frac{\hbar^2}{2m} \frac{\partial^2\Psi}{\partial \,q'^2} + \frac{1}{2} m \Omega^2 \Psi = i \hbar \frac{\partial \Psi}{\partial \tau}.
\end{equation}

For discrete solutions, we quote the results of Dantas et al. \cite{Dantas92}, but using a re-scaled variable ($a\to \sqrt{\frac{\hbar}{2m}}a$) from the ones used in their work. Their discrete solutions to the time-dependent Schr\"{o}dinger's equation are:
\begin{equation}
\begin{split}
\Psi_n(q',\tau) = (2\pi (n! \, 2^n a)^2 )^{-\frac{1}{4}}\exp{\bigg[ -\Big(\frac{q'}{2a}\Big)^2 \bigg]}\\
\times \exp{\bigg\{- i \Big( n + \frac{1}{2}\Big) \int^\tau_0 \ud t' \frac{\hbar}{2ma^2} \bigg\}} \\
\times \exp{\bigg\{  \frac{i}{\hbar}\bigg[  \frac{m \dot{a }}{2 a}q'^2 \bigg]  \bigg\}}\, H_n\Big( \frac{q'}{\sqrt{2}a}\Big),\label{ExcitedStates}
\end{split}
\end{equation}
where $H_n$ are Hermite polynomials and $a(\tau)$ is a solution to the following nonlinear equation: 
\begin{equation}
\ddot{a} + \Omega^2 a = \frac{\hbar^2}{4 m^2 a^3}. \label{aOmega}
\end{equation}

For the squeezed state solutions, we follow the approach given by Nassar \cite{Nassar02}. For more information on the method of explicitly time-dependent invariants, see \cite{LewisR,HartleyRay,Dantas92}. Nassar's full quantum squeezed state is:
\begin{equation}
\begin{split}
\Psi(q',\tau) = (2\pi a^2 )^{-\frac{1}{4}}\exp{\bigg[ -\bigg(\frac{q' - X }{2a}\bigg)^2 \bigg]}\\
\times \exp{\bigg\{ \frac{i}{\hbar} \int^\tau_0 \ud t' \bigg[ \frac{1}{2}m \dot{X}^2 - \frac{1}{2} m \Omega^2 X^2 - \frac{\hbar^2}{4ma^2}\bigg] \bigg\}}\\
\times \exp{\bigg\{  \frac{i}{\hbar}\bigg[  \frac{m \dot{a }}{2 a}\big( q' - X\big)^2 + m \dot{X}\big( q' - X \big) \bigg]  \bigg\}},\label{SqueezedStates}
\end{split}
\end{equation}
where $a(\tau)$ is a solution to Eq.~(\ref{aOmega}) as before and corresponds to the width of the wave packet; and $X(\tau)$ (which corresponds to the position of the wave packet) is a solution to the following homogenous equation:
\begin{equation}
\ddot{X} + \Omega^2 X = 0. \label{XOmega}
\end{equation}
Davydov \cite{Davydov11} showed that one can always write down a solution of Eq.~(\ref{aOmega}) given two independent solutions of Eq.~(\ref{XOmega}),  by viewing the two independent solutions as cartesian $(x,y)$-coordinates of a particle, and then realizing that Eq.~(\ref{aOmega}) is satisfied by the amplitude component $a = \sqrt{x^2+y^2}$ when we transform into polar coordinates. Figure 1 shows a plot of $|\Psi|$ for the case $X = \alpha_1\big|_{k=1,\tau_1=0}$ and $a = \sqrt{\alpha_1{}^2+\alpha_2{}^2}\big|_{k=1,\tau_1=0}$. The position of the peak starts from large positive $q'$ at a finite $\tau$ in the past ($\tau < 0$) and swings by the origin at $\tau=0$, moving off to large positive $q'$ again at $\tau>0$. During this time, the width of the Gaussian peak narrows to a sharp peak at $\tau=0$ and broadens out again. 

Here we note that the squeezed states of Nassar Eq.~(\ref{SqueezedStates}) at $X\to0$ correspond exactly to the discrete solution states of Dantas et al. Eq.~(\ref{ExcitedStates}) at $n\to0$.

\begin{figure}
\includegraphics[width=0.45\textwidth]{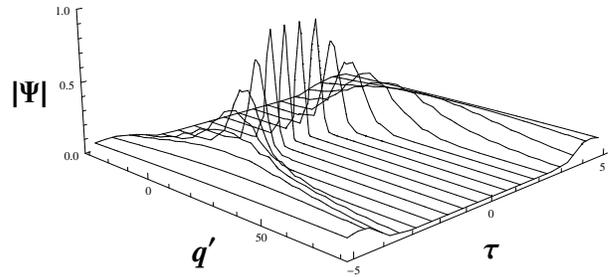}
\caption{Plot of $|\Psi(q',\tau)|$ for $X = \alpha_1\big|_{k=1,\tau_1=0}$ and $a = \sqrt{\alpha_1{}^2+\alpha_2{}^2}\big|_{k=1,\tau_1=0}$}
\label{}
\end{figure}

\section{$\beta_-'$ Hamiltonian}
Write the $\beta_-'$ Hamiltonian from Eq.~(\ref{Hbreakdown}):
\begin{equation}
H_B''=\rho \,\big(p_-'\big)^2+ 24\lambda  \ue^{4\alpha}\big(2 \ue^{4\beta_+}-\ue^{-2\beta_+}\big)\big(\beta_-'\big)^2
\end{equation}
Here, we cannot use `differentiation with respect to parameters' to generate explicit forms of the linearized variables. This is because the perturbations do not stay within the Taub family. However, all the results from above still apply to this hamiltonian. Specifically, one can imagine the same set of states that are labeled by an independent integer.

\section{Discussion}
In our previous work, we demonstrated the application of the modified semi-classical method to the canonically quantized vacuum Bianchi IX (Mixmaster) model, solving the relevant Wheeler-DeWitt equation asymptotically by integrating a set of linear transport equations along the flow of the Moncrief-Ryan (or `wormhole') solution to the corresponding Euclidean-signature Hamilton-Jacobi equation \cite{Bae2014}. We found that the excited state solutions, peaked away from the minisuperspace origin, are labeled by a pair of positive integers that can be plausibly interpreted as graviton excitation numbers for the two independent anisotropy degrees of freedom $(\beta_+, \beta_-)$. Is there some way to help us interpret these states in a more intuitive way?

In this paper, we have taken a different approach to the approximate solutions of the Bianchi IX model, using the Jacobi method of second variation to study the homogeneous linearized perturbations about the Taub background solution. 

By way of visualization afforded by the explicit quantum squeezed states of the linearized Hamiltonian for $Q'=p_+ \alpha' + p_\alpha \beta_+'$, we have sought to give a more intuitive explanation for the excited states found in our previous work.

In conclusion, we have applied the method of invariants to a linearized perturbation model about the exact Taub background. Our methods used here apply to a wider set of models used in quantum cosmology. The Jacobi method of second variation can also be extended to inhomogeneous perturbations about any background model formulated in the variational principle.




\section*{Acknowledgements}

I am grateful to Professor V. Moncrief for suggesting this problem and for his advice and encouragement throughout the course of this work. I would like to thank Yale University for financial support.

\bibliography{JHBPHD}{}
\bibliographystyle{ieeetr}

\end{document}